\documentstyle [12pt]{article}
\renewcommand{\l}{\ell} %spaziatura
\newcommand{\be}{\begin{equation}}
\newcommand{\ee}{\end{equation}}
\newcommand{\bdis}{\begin{displaymath}}
\newcommand{\edis}{\end{displaymath}}
\newcommand{\eg}{\varepsilon}
\newcommand{\ep}{\epsilon}

\title{Chaotic
cascades with Kolmogorov 1941
scaling}

\author{L. Biferale$^{1}$, M. Blank$^{1,2}$ and U. Frisch$^{1}$}

\begin{document}
\maketitle
\centerline{$^{1}$
CNRS URA 1362, Observatoire de Nice, BP 229, 06304 Nice Cedex 4, France.}
\centerline{$^{2}$
Russian Academy of Sci., Inst. Gen. Genetics, Gubkin st. 3,
 117809, Moscow, Russia.}
\date{ }
\medskip
\centerline{\it Submitted to J. Stat. Phys.}

\begin{abstract}

We define a (chaotic) deterministic variant of random multiplicative
cascade models of turbulence. It preserves the hierarchical tree structure,
thanks to the addition of infinitesimal noise. The zero-noise limit
can be handled by Perron-Frobenius theory, just as the zero-diffusivity limit
for the fast dynamo problem. Random multiplicative models do not
possess
Kolmogorov 1941 (K41) scaling because of a large-deviations effect.
Our numerical studies indicate that {\it deterministic}
multiplicative models
can  be chaotic and still
have exact K41 scaling. A mechanism is
suggested for avoiding large deviations, which is present
in  maps with a neutrally
unstable fixed point.

\end{abstract}

\vspace{2.truecm}
\newpage

\section{Introduction}

One  popular way to describe the small-scale activity of fully developed
turbulence is to suppose that energy is transferred from the injection
scale to
the viscous scales, by a multi-step process along the inertial range.
This  idea has been often used to predict
important features of turbulent flows. Still, relations with the structure of
Navier-Stokes equations are  poorly understood.

In his 1941 work, Kolmogorov \cite{k41}
 uses his phenomenology to postulate that all the
statistical properties of the turbulent flow at scales belonging to the
inertial range are {\it universal}, in the sense that they depend only on the
scale $\l$ and on the mean energy dissipation rate per unit mass $\bar{\eg}$.
As a consequence, moments of velocity increments, over small distances $\l$,
should posses universal forms. This led Kolmogorov to
dimensionally-based expressions for the structure functions\,:

\be
S_{\l}^{(p)} \equiv E\{ \delta v_{\l}^p(x)\} \equiv
E\{(v(x+\l)-v(x))^p\} \sim
(\bar{\eg}\l)^{p/3},
\label{eq:sca}
\ee

\noindent where $E\{ \cdots \}$
denotes ensemble average and the symbol $\sim$ means equality within $O(1)$
multiplicative constants. Thus
(\ref{eq:sca}) predicts scaling behavior for the structure functions,
 the function of
order $p$ having the exponent $\zeta_p=p/3$.

 Although early
experimental data seemed to confirm  the prediction
 (at least for $p=2$),
the K41 theory has been  criticized because it does not take into account the
natural (and also experimentally verified) presence of
 fluctuations in the energy dissipation.
 These fluctuations are commonly believed to be the
consequence of the chaotic transfer of excitations along the inertial range.
A possible way to account for  this effect, suggested by Obukhov \cite{ok},
 is the following. First, one  introduces the average rate of
energy dissipation over a cubic
box $\Lambda_{\l}(x)$ of side $\l$ centered on $x$\,:

\be
\ep_{\l}(x) = \frac{1}{\l^3}\int_{\Lambda_{\l}(x)} \eg(x)dx.
\label{eq:ene}
\ee
Second,  fluctuations in the velocity increment are related
to fluctuations in $\ep_{\l}(x)$ by a ``bridging relation''
suggested by Kolmogorov's 1941 theory, {\it viz}
\be
\delta v_{\l}^p(x) \sim (\ep_{\l}(x))^{1/3}.
\label{eq:bridge}
\ee
This bridging relation is still widely used. It has received good
 experimental support \cite{mevsre}. Still it leads to some conceptual
difficulties which are unrelated to the material to be presented
in this paper \cite{aurell}. From (\ref{eq:ene}) and (\ref{eq:bridge}),
we find
\be
S_{\l}^{(p)} \sim E\{{\ep_{\l}}^{p/3}\}\,\l^{p/3}.
\label{eq:scamulti}
\ee

Considerable experimental \cite{ans,sre,vm} and theoretical
work \cite{parfri,bppv,mevsre}
 has been devoted in recent years to measuring and
predicting the dependence of the quantities
$E\{\ep_{\l}^{p/3}\}$ on the scale $\l$.
To date, it seems that a certain consensus
 has been reached. From the experimental side, there is clear
evidence of a non-trivial
$\l$-dependency (at least for high order moments). Similarly,
from a theoretical point of view, all models based on {\it random}
multiplicative  cascades (Section 2)
 introduce power-law corrections to K41 scaling.  The fact that
 these simple
models  have deviations to K41 scaling has led perhaps
to the misconception that any  cascade model having
non-trivial fluctuations is inconsistent
 with K41 scaling.

Actually, we shall show
 that the chaotic transfer of energy described by  a
deterministic multiplicative process can still be  consistent with K41,
or more precisely that possible deviations disappear in the ``fully
developed limit'', i.e. for cascades with very many steps.

The paper is organized as follows. In Section 2, we shall
 connect the well-known case of random
independent multiplicative cascade models and our (deterministic) case
 in two
steps. First, we introduce a  Markov random model and then, by means of a
special limiting construction, we obtain the deterministic model.
A particular class of deterministic multiplicative models
is introduced in Section 3 and studied numerically
in Section 4.
Finally,
in Section 5 we give a possible interpretation of the lack of corrections
to K41 scaling observed in our model.

\section{Pure-random and noisy-deterministic multiplicative models}

Random multiplicative models were introduced by Novikov and Stewart \cite{ns}
and by Yaglom \cite {yag} as a simple way to
describe stochastic transfer of energy along the inertial range.
Their fractal properties were discussed by Mandelbrot \cite{man}.
Let us give now the definition of these models. A binary
 tree structure, obtained by hierarchically  partitioning the
original volume of size $\l_0$ in subvolumes of size $\l_n=2^{-n}\l_0$,
is used
to describe fluctuations at different scales (fig. 1a illustrates
 the  one-dimensional case).
The energy dissipation, $\ep_n$, associated to a cube at scale
$\l_n$, is multiplicatively
 linked to the energy dissipation, $\ep_{n-1}$, at the
larger scale, $\l_{n-1}$, through a random variable
$W_n$\,:

\be
\ep_n=W_n \ep_{n-1} = W_nW_{n-1}W_{n-2}....W_1 \bar{\eg}.
\label{eq:multi}
\ee
The $W_n$'s are identically and independently distributed positive
random variables. The structure functions
are now defined as
\be
S_n^{(p)} \equiv E\{\ep_n^{p/3}\}{\l}_n^{p/3}
\label{eq:sn}
\ee
Using (\ref{eq:scamulti}) and (\ref{eq:multi}),
we obtain\,:
\be
S_n^{(p)}
=  \l_n^{\zeta(p)},\qquad {\rm with} \qquad \zeta(p)=p/3-
\log_2 E\{W^{p/3}\},
\label{eq:multi2}
\ee
\noindent where $E\{\cdots\}$ denotes the mathematical expectation.
Here, $-\log_2 E\{W^{p/3}\}$ is the correction to the K41 exponent in
the structure function of order $p$.
Actually, the multiplicative model (\ref{eq:multi}) has very interesting
properties when considered in the light of the theory of large
deviations,  as noted, in particular
by Oono \cite{ono}
 and Collet and Kukiou  \cite{coku}.
It follows from Cram\'er's \cite{cra} (see also Refs.~\cite{vara}
and \cite{ell})
 work that, for large $n$'s, the
 quantity\,:

\be
\frac{1}{n}\log_2 \ep_n =
\frac{1}{n}(\log_2 W_1+\log_2 W_2 + \cdots + \log_2 W_n)
\label{eq:ld}
\ee
can deviate from its limit value, $E\{\log_2 W\}$, by a finite non-vanishing
amount $\alpha$ with a probability which decreases as $\exp(-f(\alpha)n)$.
 The
 Cram\'er\footnote{A name suggested by Mandelbrot \cite{man91} }
function $f(\alpha)$ is positive and convex and
can be identified with an entropy in applications to
statistical thermodynamics \cite{ell}.
 The presence of such  large deviations, with probability decreasing
exponentially in $n$ (i.e. as a power law in $\l_n$),
introduces fluctuations in the effective scaling exponent and therefore an
overall change of the scaling properties of all structure functions. Indeed,
we can rewrite (\ref{eq:multi2}) in the following form\,:

\be S_n^{(p)} \propto E\{\ep_n^{p/3}\}\l_n^{p/3} =
E\left\{\l_n^{-\frac{p}{3} \frac{1}{n} (\log_2 W_1+\log_2 W_2 + \cdots +
 \log_2
W_n)}\right\}\l_n^{p/3} \propto  \l_n^{\zeta_p}. \ee

\noindent  As is well-known, in large-deviations theory,
 the function $\zeta(p)$ is given by a Legendre
transformation\,:
\be
\zeta(p)= \inf_{\alpha}\left(\frac{p}{3}(1-\alpha) +f(\alpha)/\ln 2 \right).
\ee

Our aim now is to construct a deterministic variant of such   models. We
shall do it in two steps. First, we consider, instead of independent
random variables $\{W_i\}$, successive  points on the orbit of a Markov
process. This means
that we consider a Markov process on a phase space $X$ with a transition
probability operator $P$ and an observable $h: X \to R_+^1$. The
(generalized)
structure
function at the scale $\l_n=2^{-n}\l_0$
is then  defined as\,:

\be
S_n^{(p)} =  S_n^{(p)}[h,P] = E\left\{\prod_{k=1}^n h^p(x_k)\right\},
\label{eq:ppp}
\ee

\noindent where $x_k \in X$ are points of an orbit of the  Markov
process. Introducing now\,:
\be
 Z_n^{(p)}=\prod_{k=1}^n h^p(x_k), \ee we easily find
 that the conditional  expectation of $Z_n(p)$ with
respect to the initial distribution density, $\rho(x)$, is
given by\,:
\be
E\{Z_n^{(p)}| \rho \} = \int_X h^p(x)
\rho(x) P_{h,p}^n \,1(x) \, dx .
\label{eq:markov}
\ee

\noindent Here, the operator $P_{h,p}$ is defined by
the relation $P_{h,p} \phi(x) =
h^p(x)P\phi(x)$.
This concludes the first step. The second step is to change from
a  Markov {\it random} dependence to a
{\it deterministic} dependence by use of a chaotic map.
It seems that this can be done in an obvious way, replacing
the transition probability operator by the
transfer-matrix (Perron-Frobenius operator) of the chaotic map. However, some
additional work is needed to define a deterministic construction
possessing the same tree structure as in the random case (fig. 1a).
Indeed, with a deterministic map, how can we avoid
giving the same value to the two off-springs of the next generation,
thereby trivializing the whole tree-structure?

It is necessary to define a consistent procedure to distinguish the
branches of
the tree. We build up this branching-deterministic process by inserting a
 small
amount of noise at each node and by considering the total process as the
superposition of the deterministic transfer along
consecutive levels plus the
noise. The final  map will be obtained by taking the zero-noise limit.

Let $f:X \to X \in R^d$ be the  deterministic map which describes the
relation
between  any two consecutive scales of the tree structure (see fig. 1b), and
let us fix an observable $h:X \to R^1_+$. Then for any fixed number
$\eta > 0$
we may consider a random Markov chain $(x_k^{(\eta)})$ such that

\be
 x^{(\eta)}_{k+1} = f(x_k^{(\eta)}) + \eta \,\xi_{k},
\label{eq:mapmarkov}
\ee

\noindent where the $\xi_{k}$'s are random variables, independently
and identically distributed on the interval
$[-1,1]$. Now, we may apply the previous construction by
considering different realizations of the noise $\xi_{k}$ on each link
connecting consecutive nodes. Then, any cascade process along a branch of the
tree in fig. 1b will be described by the map (\ref{eq:mapmarkov}) with
different realizations of the noise.

We define our structure functions as the zero-noise limit
of (\ref{eq:ppp}).

\noindent
For a fixed value of $\eta > 0$ it follows from  (\ref{eq:markov})
 that
the large-$n$ limit of the structure functions is governed by the
spectrum of the  operators\,:

\be  P_{h, p, \eta} = h^p Q_\eta P.
\label{eq:truc}
\ee

\noindent Here, $Q_\eta$ is the transition operator for the
 random perturbation,
$P$ is the transfer-matrix, or Perron-Frobenius operator
 of the map $f$ and $h$ is an observable (for definitions see
for example Ref.~\cite{B1}).

Let us mention, incidentally, that there is also another possibility
to define a deterministic (chaotic) process on a tree structure.
Using a method of finite-state Markov approximations of chaotic
maps, proposed by Ulam, we obtain a Markov chain
with transition probabilities given by
$ p_{ij} = \vert X_i \cup f^{-1}X_j \vert / \vert X_i \vert$,
where the set of $X_i$'s defines a suitable finite partition
of the original phase space $X$. It is known, for instance, that
for sufficiently ``good'' maps (for example piecewise expanding with
derivatives larger than two), invariant distributions of this simple
Markov chain converge to the density of the invariant measure of
the map under consideration when $\vert X_i \vert \to 0$.
With this Markov process, we construct the analog of (\ref{eq:truc}),
{\it viz}
 the operator $P_{h, p}=h^pP$. (Here $P$ is the
transition operator with matrix elements $p_{ij}$.)
These possibilities will not be explored further here.\\

We now observe that the construction based
on taking the zero-noise limit
appearing in (\ref{eq:truc}) is
  similar  to that of the mathematical theory of {\it
fast dynamos} \cite{arno,gilbe,O}.
Fast dynamo theory
describes the phenomenon by which rapid
 magnetic field growth can be sustained in the presence of a
 prescribed velocity
field, when taking the zero-diffusivity limit.
{}From a formal point of view,  fast dynamo theory involves a combination
of two operators\,: a transfer-matrix for some deterministic
map, associated to a deterministic velocity field,
 and a diffusion-like operator, or equivalently small-amplitude noise.

The main purpose  of
the theory is then to find the
 properties of the zero-noise limit of the combined
operator.
Oseledets \cite{O}, in his study of the
dynamo problem, considered a noise-perturbed Perron-Frobenius operator
similar to that of (\ref{eq:truc}) and was able to show
in special cases that it converges for $\eta \rightarrow 0$
to the  Perron-Frobenius operator for the deterministic problem. The latter
is than said to be ``stochastically stable''.\\

Returning to the structure function of the multiplicative model, we shall also
assume that the Perron-Frobenius operator is stochastically stable. In
addition, we assume that the map (\ref{eq:mapmarkov}),
with $\eta = 0$, is chaotic and ergodic. We can then compute the
structure functions as averages along-the-orbit for the
deterministic map, i.e.
\be
S_n^{p}=\lim_{N \rightarrow \infty} \frac{1}{N} \sum_{i=1}^N
\prod_{k=1}^n h^p(x_{i+k})
\label{eq:mostro}
\ee
with $x_{i+1}=f(x_i)$.
\label{sec:2}

\section{ A class of deterministic cascade models based on shell models}

The construction defined in Sec~\ref{sec:2}, after taking the zero-noise
limit as explained, amounts simply to the following\,:
 we keep (\ref{eq:multi})
as it stands, but instead of searching the $W_n$'s randomly,
 we assume that\,:
\be
W_n = g\left( W_{n-1},W_{n-2},....,W_{n-r}\right),
\label{eq:bb}
\ee
where $g$ is a deterministic map which involves a finite number
$r$ of antecedents.

There is an alternative formulation in terms of
characteristic velocities associated
to the hierarchy of scales $l_n = l_0 2^{-n}$. By analogy with
(\ref{eq:bridge}), we introduce a set of velocity variables, denoted $u_n$,
related to the $\ep_n$'s by\,:
\be
\
\ep_n =\frac{|u_n|^3}{l_n}.
\ee
The $u_n$'s can be thought of as  (real or complex)
velocity amplitudes associated to eddy-motion on scale $l_n$. Instead of
(\ref{eq:multi}) we then use
\be
u_n = q_n\,q_{n-1}\,q_{n-2}....q_{1}\,u_0,
\ee
while (\ref{eq:bb}) becomes\,:
\be
q_n=f(q_{n-1},.....,q_{n-r}).
\label{eq:mm}
\ee
Eq.~(\ref{eq:mm}) will be called the ``ratio-map'' since the $q_n$'s are the
ratios of
successive velocity amplitudes.

We shall now consider a particular class of ratio maps generated
from ``shell models''. The latter can be viewed as the poor man's
Navier-Stokes
equations\,: instead of the whole velocity field, one retains
 only a discrete set
of velocity amplitudes $u_n$, the dynamics of which is governed
by a set of coupled differential equations of the form\,:
\be
\dot{u}_n = F_n(u_{n-s},...,u_n,...,u_{n+s}),
\ee
where the $F_n$'s involve finitely many neighbors (in scale) of $u_n$. The
various $n$'s are referred to as ``shells''. Except for viscous and forcing
terms, the $F_n$'s are chosen quadratic and satisfy energy conservation.
 We shall not here attempt any review of the considerable literature
 on shell
models (see e.g. Refs.~\cite{novdes,yamohk,jpv,pbcfv}, and references
 therein).
Our intention is to use a {\it static} form of shell models, i.e. we
assume\,:
\be
F_n(u_{n-s},...,u_n,...,u_{n+s})=0.
\ee
It is not our intention to discuss such
thorny issues as\,: are shell models ``good'' approximations to true
turbulence? Can we learn something about the dynamics of shell models by
studying the time-independent
solutions?

We just observe that static shell models immediately generate
ratio-maps of
the form (\ref{eq:mm}). Indeed, ignoring viscous and forcing terms, which
are not relevant in the inertial range, and assuming that
 $F_n$ is a homogeneous polynomial
of degree two in the $u_n$'s, we find that $q_n=u_n/u_{n-1}$  satisfies
a recursion relation
 of the form (\ref{eq:mm}). An example will make this clear. One
popular model is the Gledzer-Ohkitani-Yamada (GOY) model
\cite{gle,yamohk} which, in its complete form (with viscosity, forcing
and time-dependence), reads\,:
\be
\dot{u}_n^* +\nu  k_n^2u_n^* +f_n=F_n \equiv
 -ik_n\left( u_{n+1}u_{n+2}-\frac{1}{4}
u_{n+1}u_{n-1}-\frac{1}{8}u_{n-2}u_{n-1} \right).
\label{eq:shell}
\ee
Here, $\nu$ is the viscosity,  $f_n$ is the forcing, and $k_n=l_n^{-1}$ is
the shell-wavenumber. The inviscid, unforced and static GOY model gives\,:

\be
 u_{n+1}u_{n+2}-\frac{1}{4}
u_{n+1}u_{n-1}-\frac{1}{8}u_{n-2}u_{n-1} = 0.
\ee
Hence, the recursion relation for successive ratios is
\be
q_{n}=\frac{1}{q_{n-1}q_{n-2}}
 \left[ \frac{1}{4}+\frac{1}{8}\frac{1}{q_{n-1}q_{n-2}q_{n-3}} \right].
\label{eq:map2}
\ee
It turns out that the ratio-map generated by the GOY model is
quite trivial. Indeed, from (\ref{eq:map2}) it follows that the
product $z_n=q_n\,q_{n-1}\,q_{n-2}$ satisfies a first order recurrence
relation which has a single stable fixed point $z_n=z_*=2^{-1}$.
Hence, there is no chaos, and K41 scaling is obtained.
 To obtain chaotic
variations (in the shell index), we need to have more terms
in $F_n$. One simple way is to perturb the GOY model by putting an
 admixture of another popular shell model, the
 Desnyanky-Novikov (DN) model \cite{novdes}.
In this way we generate the following ratio map\,:

\begin{eqnarray}
q_n & = &f(q_{n-1},q_{n-2},q_{n-3}), \nonumber \\
f(x,y,z) & = &
\frac{1}{xy} \left( \left( \frac{1}{4}+\frac{1}{8}\frac{1}{xyz} \right)
+ \gamma \left(\frac{1}{xy} -2y +\delta(\frac{1}{x}-2xy) \right) \right).
\label{eq:map}
\end{eqnarray}
Here, $\delta$ is the so-called ratio of  backward to forward cascade
amplitudes
in the DN model and $\gamma$ is the admixture parameter, which in
subsequent numerical computations will typically be taken small
(about $10^{-2}$).
The hybrid ratio-map (\ref{eq:map}), which will be referred to as GOY-DN
in the sequel, has still a K41 fixed point $q_n=q_* =2^{-1/3}$, but
it  is easy to find a region in
the  phase-space of the parameters $\gamma$ and $\delta$
where the K41 fixed point is unstable; for example
 $\gamma > 0, \, \forall \delta$.
Further information about the behaviour of (\ref{eq:map}) requires
numerical simulations.

\section{Numerical results}

 We found numerically that the GOY-DN map,
defined by (\ref{eq:map}) develops  chaotic behavior for
a sizable set of values of the control parameters
 $\gamma$ and $\delta$.
Most of the  numerical simulations reported hereafter were done with
$\gamma=0.01$ and $ \delta=4$. However, the qualitative feature of the results
appear to be very stable against changes in the
parameters, as long as the dynamics are chaotics.

Fig. 2a  shows a typical orbit of the map (\ref{eq:map}). The initial
condition is obtained by adding a very small random perturbation
(r.m.s. value about $10^{-5}$) to the K41 fixed point value. The orbit looks
chaotic and also displays strong intermittent fluctuations.
 Its largest Lyapunov
exponent is found to be about $0.22$.
The Kolmogorov fixed point, although it ceases
to be stable as soon as $\gamma > 0$,   still plays an
important role in the dynamics. Fig. 2b shows an enlargement of fig 2a,
revealing
 that the orbit mostly oscillates
 close
to the K41 fixed point,  occasionally going on  wild excursions.
 We  mention that in the
{\it unperturbed } GOY map, the K41 fixed point is neutrally stable\,: two
eigenvalues of the derivatives matrix, calculated at the fixed point,
have modulus  one. Of course, the fixed point becomes
completely unstable as soon as $\gamma >0$. \\

We now turn to the structure functions (\ref{eq:mostro}),
evaluated by averaging along the orbit with several million points.
Fig. 3 shows the structure functions for $p$ from one to four. They are
seen to follow power-laws (in $k_n$) at large $n$'s. Except for some
residual chaotic noise, the exponents are exactly given by their K41 value\,:
 $\zeta(p)=p/3$.  Here, a word of
warning is required. In numerical simulations
of the full time-dependent shell models, it is not practical to have
more than, say, 30 shells, since time steps
decrease exponentially with $n$. Had we used such a low value of $n$ for
estimating
exponents of structure functions, we would have predicted erroneous
values, which would be misread as ``multifractal'' corrections to K41.
We do not want here
to open the Pandora-box of whether the
multifractal corrections detected in the time-dependent
 simulations of shell models
\cite{jpv,pbcfv} are or are not   finite-shell artefacts (similar
questions can be asked for finite-Reynolds number
turbulence data).\\

The presence of exact K41 scaling suggests that the GOY-DN model has no
large deviations.
This can be checked directly by plotting the p.d.f. of the normalized
sums of logarithms\,:
\be
\Sigma(n)=\frac{1}{n}\sum_{k=1}^n \log_2 |q_k|,
\label{eq:28}
\ee
playing in the deterministic case the role of
 (\ref{eq:ld}) in the random case.

Fig. 4  shows this p.d.f. for two values of $n$. Note that
as $n$ grows it is increasingly peaked at the K41 value   $-1/3$.
It is thus likely that for $n \rightarrow \infty$,
it becomes a delta function at this value. A possible interpretation of
this phenomenom is presented in the next Section.

\section{Neutrally unstable and sporadic maps}

We present a very simple example of  a class of maps which displays chaos
and no large  deviations. Let us define\,:
\be
\label{eq:frac}
f_z(x) = {\rm Frac}\,(x^z+x): \, [0,1] \to [0,1] \,\,,\,\,z \ge 2.
\ee
The deterministic maps are chaotic,
 piecewise expanding, and have  a neutral (or indifferent)
fixed
point at the origin. These maps are  ``sporadic'', in the following  sense
\cite{gw}\,: The growth of infinitesimal errors is controlled by
a stretched exponential with exponent less than one. Hence, the Lyapunov
exponent is zero, although the maps certainly deserve to be called ``chaotic''.
 The
derivative of any of these  maps is equal to one
at the origin and is strictly larger than one at all other points where it is
well-defined. This leads to some unusual
ergodic properties, for example the nonintegrability of the
invariant measure, the density of which
 has a power-law singularity
near the origin \cite{B2,CF,gw}.
The reason is that  once the orbit gets  very
close to the neutral fixed point, the
time to escape from it is  inversely proportional to a power of
the distance
from the fixed
point.
 As a consequence most usual ergodic averages such as (\ref{eq:28})
 will be dominated by the
contribution from the  fixed point, so that there are no
large deviations.

As for the absence
of large deviations in the GOY-DN map (\ref{eq:map}) we are led,
at this moment, to speculate that something similar happens\,:
the presence of a neutrally unstable
fixed point (or, more likely, a neutrally unstable periodic orbit)
dominates the large-$n$ asymptotics  of
structure functions.

Let us finally make two  remarks. First,
the positivity of the largest Lyapunov exponent in
the GOY-DN map (\ref{eq:map})
is not inconsistent, as one would think at first,
 with the kind of
 behavior displayed by the  maps (\ref{eq:frac}) which have zero Lyapunov
exponent. To illustrate this, consider
 the direct product of two
one-dimensional maps\,: the first is
any  map chosen in the class  (\ref{eq:frac})
 and the second is the piecewise linear map $g(y) = {\rm
Frac}(2y)$ from $[0,1]$ to itself. The product
has as its  largest Lyapunov exponent  $\log
2$. Nevertheless,
 for the observable $h(x,y)=xy$, ergodic averages are still dominated by
the contribution from $x=0$,
as if the multiplier $y$ was absent. Second, let us mention
that for the  maps
(\ref{eq:frac}) it is
possible to give  an estimate of the average fraction of time during which
the dynamics is chaotic \cite{gw}. This average goes to zero as a power-law of
the number of iterations $n$  (e.g.  as $\log n/n$ when $z=2$). If
the analogy between the sporadic maps and the GOY-DN model is
correct, this  suggests  a possible interpretation
 of the rather strong  corrections
to  K41 scaling at {\it finite} $n$, which eventually disappear
at large $n$.

\vspace{1.truecm}

\par\noindent{\bf ACKNOWLEDGMENTS}

The authors would like to thank V. Oseledets for discussions on the
fast dynamo
problems and G. Paladin for discussions
on sporadicity and shell models. L. B. was supported by a Henri Poincar\'e
fellowship
 (Centre National de la Recherche Scientifique and Conseil
G\'en\'eral des Alpes Maritimes). M. B. was supported by the
 French Ministry of Higer
Education.

\newpage

\centerline{{\bf Figure Captions}}

\noindent {\bf Figure 1}\,: a) A one-dimensional representation of the dyadic
tree for the random multiplicative chaos model. Each variable $\{W_i\}$ is
chosen independently and identically distributed. b) The branching process for
the noisy deterministic maps; the  $\xi$'s  are chosen independently
on each link.  \vspace{0.5truecm}

\noindent {\bf Figure 2}\,: a) A typical solution of the difference equation
(\ref{eq:map}). b) A zoom of the previous figure. The horizontal solid line
corresponds to the Kolmogorov fixed point.

\vspace{0.5truecm}

\noindent {\bf Figure 3}\,: Log-plot of the structure
functions $S_n^{(p)}$ as a
functions of $n$, for $p=1,2,3,4$. The solid lines correspond to the
Kolmogorov
1941 scaling.

\vspace{0.5truecm}

\noindent {\bf Figure 4}\,: Probability density functions (p.d.f.) of
$\Sigma(n)$, defined by (\ref{eq:28})
 for $n=21$ and $n=36$. Notice that the peak at the Kolmogorov value,
$\Sigma(n)=-1/3$ becomes more pronounced as $n$ increases.


\vspace{1.5truecm}


\begin{thebibliography}{99}

\bibitem{k41} A. N. Kolmogorov, {\it Dokl. Akad. Nauk SSSR} {\bf 30}, 301
(1941).

\bibitem{ok}A. M. Obukhov, {\it J. Fluid. Mech.} {\bf 13} 77 (1962).
\bibitem{mevsre} C. Meneveau and K.R. Sreenivasan, {\em J. Fluid Mech. }
 {\bf
224}, 429 (1991).

\bibitem{aurell} E. Aurell, U. Frisch, J. Lutsko and M. Vergassola,
{\it J. Fluid Mech.} {\bf 238}, 467 (1992).

\bibitem{ans} F. Anselmet,  Y. Gagne,  E.J.  Hopfinger and  R.A. Antonia,
 {\it
J. Fluid Mech.} {\bf140},  63 (1984).

\bibitem{sre} C. Meneveau and K.R. Sreenivasan, {\it Nucl. Phys} {\bf B}
 [Proc.
Suppl.], {\bf 2}, 49 (1987).

\bibitem{vm} A. Vincent and  M. Meneguzzi, {\it J. Fluid Mech.} {\bf 225}, 1
(1991).

\bibitem{parfri} G. Parisi and  U. Frisch, in {\it Turbulence and
Predictability in Geophysical Fluid Dynamics}, Proceed. Intern. School of
Physics ``E. Fermi'', 1983, Varenna, Italy, eds. Ghil M., Benzi R. and Parisi
G., p. 84, North-Holland (1985).

\bibitem{bppv} R. Benzi, G. Paladin, G. Parisi and A. Vulpiani, {\it J. Phys.}
{\bf A 17}, 3521, (1984).



\bibitem{novdes} V.N.  Desnyansky and  E.A. Novikov,
{\it Izv. Akad. Nauk SSSR
Fiz. Atmos. Okeana} {\bf10},  127 (1974).

\bibitem{gle} E.B. Gledzer, {\it Sov. Phys. Dokl.} {\bf 18},  216 (1973).

\bibitem{yamohk} M. Yamada and K. Ohkitani, {\it J. Phys. Soc. Jap. }
 {\bf 56},
4210 (1987).
\bibitem{ns}
 E.A. Novikov and  R.W. Stewart
 {\it Izv, Akad. Nauk SSSR, ser. geoffiz.} {\bf 3},
408 (1964).
\bibitem{yag} A. Yaglom, {\it Dokl. Akad. Nauk. SSSR} {\bf 166}, 49 (1966).

\bibitem{man} B.B. Mandelbrot  {\it J. Fluid Mech.} {\bf 62}, 331  (1974).

\bibitem{ono}
 Y. Oono {\it Progr. Theor. Phys. Suppl.} {\bf 99},
165  (1989).
\bibitem{coku}
 P. Collet  and   F. Koukiou
 {\it Commun. Math.
Phys.} {\bf47}, 329 (1992).
\bibitem{cra} H. Cram\'er ``Actualit\'es scientifiques et industrielles''
{\bf 736},  5; Colloque consacr\'e \`a la th\'eorie des probabilit\'es,
Hermann, Paris 1938.
\bibitem{vara}
S.R.S. Varadhan  ``Large Deviations
and Applications'', SIAM (1984).



\bibitem{ell} R.S. Ellis  ``Entropy and Equilibrium States in Classical
Mechanics'' Springer (1985).

\bibitem{man91}  B.B. Mandelbrot
{\it Proc. R. Soc. Lond. A} {\bf 434},  79 (1991).

\bibitem{B1} M. Blank, {\it Uspekhi Matem. Nauk.} {\bf 44}, 3 (1989).

\bibitem{arno} V. I. Arnold, Ya. B. Zeldovich, A. A. Ruzmaikin
and D. D. Sokoloff {\it Sov. Phys. J.E.T.P.}
{\bf 54}, 1083 (1981).
\bibitem{gilbe} E. Aurell and A. D. Gilbert,
``Fast dynamo and determinants of singular integral operators''
{\it Geophys. Astrophys. Fluid Dyn.} (1993), in press.
\bibitem{O} V. Oseledets, ``Fast dynamo problem for
a smooth map on two-torus'', to be published.

\bibitem{jpv} M.H. Jensen, G. Paladin and A. Vulpiani, {\it Phys. Rev.}
{\bf
A43},
798 (1991).

\bibitem{pbcfv} D. Pisarenko, L. Biferale, D. Courvoisier, U. Frisch and M.
Vergassola, {\em Phys.  Fluids} {\bf A5}, 2533 (1993).


\bibitem{B2} M. Blank, {\it Chaos} {\bf 1}, 347 (1991).

\bibitem{CF} P. Collet and P. Ferrero, {\it Ann. Inst. H. Poincar\'e} {\bf 52},
283 (1990).

\bibitem{gw} P. Gaspard and X.-J. Wang, {\it Proc. Natl. Acad. Sci. USA}
{\bf 85}, 4591 (1988).

\end{thebibliography}
\end{document}